# Design Considerations for Building Credible Security Testbeds
A Systematic Study of Industrial Control System Use Cases

Uchenna D Ani[1], Jeremy M Watson[2], Benjamin Green[3], Barnaby Craggs[4], and Jason Nurse[5]

[1, 2]Department of Science Technology Engineering and Public Policy, University College London

[3]School of Computing and Communications, Lancaster University, UK

[4]Department of Computer Science, University of Bristol, UK.

[5]School of Computing, University of Kent, UK

*{Corresponding Author: [1]u.ani@ucl.ac.uk}*

**Abstract:**
This paper presents a mapping framework for design factors and implementation process for building credible Industrial Control Systems (ICS) security testbeds. The resilience of ICSs has become a critical concern to operators and governments following widely publicised cyber security events. The inability to apply conventional Information Technology security practice to ICSs further compounds challenges in adequately securing critical systems. To overcome these challenges, and do so without impacting live environments, testbeds for the exploration, development and evaluation of security controls are widely used. However, how a testbed is designed and its attributes, can directly impact not only its viability but also its credibility as a whole. Through a combined systematic and thematic analysis and mapping of ICS security testbed design attributes, this paper suggests that the expertise of human experimenters, design objectives, the implementation approach, architectural coverage, core characteristics, and evaluation methods; are considerations that can help establish or enhance confidence, trustworthiness and acceptance; thus, credibility of ICS security testbeds.

## 1. Introduction

Industrial Control Systems (ICSs) are essential components of critical national infrastructures (CNIs) that control societal services, e.g. power generation, water treatment, and transport infrastructure. Whilst advances in technology have improved ICS functionality through fundamental design, setup, and operational scope, the reality is much more challenging. ICSs can have real-world deployment life cycles measured in decades, leading to outdated and insecure legacy systems running alongside modern, more secure deployments. The security of ICSs has become a growing concern owing to observed challenges and real-world cyber-attacks [1]. Where ICSs form a core component of CNI, they must provide high levels of system safety, availability, security, and operational resilience. These requirements can be tied to several factors, where the impact of failure from economic, environmental, human safety, and national security perspectives would be highly detrimental [2]. In addition, testing these requirements is often impracticable in real-life ICS operational environments because of the potential disruption to process functions that can occur [3,4].

Testbed development has grown to the point where replicating ICS networks through modelling and simulation (M&S) is considered a viable alternative for exploring and addressing cybersecurity challenges [5]. This is, in part, due to the high cost of deploying and using real system hardware and software for testing purposes, and the obvious risks linked to conducting research-based tests upon live ICSs or Operational Technology (OT). In M&S, a model of an actual system or problem is used rather than directly working on the real (actual) physical system.

Emulation of ICS can be approached in various ways providing exploratory platforms upon which experimentation and training can be performed safely, avoiding socio-economic impacts associated with performance degradation [6]. This process is more technically referred to as *'Simulation'*, and is widely acknowledged to be effective in experimenting, studying, analysing, and developing ICS security solution best practices [7].

Whilst the term *'Simulation'* is commonplace, often such environments can also be referred to as *'Testbeds'*. This encompasses the range of setups that include hardware-in-the-loop (HIL) as a step between *'computer simulation'* and *'physical operational hardware simulation'*. These provide test platforms for executing activities and processes as if in a real-world environment. For clarity, and for the remainder of this paper, we will use testbeds as an all-inclusive term.

A number of ICS testbeds have been developed and reported by researchers to analyse or address security-related challenges [8]. These papers describe a range of approaches to the research [7]. A key challenge is to model and simulate real control system conditions accurately and with enough detail to support confidence in the simulated system together with its attributes and processes, thus enhancing trustworthiness and acceptability. Confidence depicts a state of certainty, either about the correctness of a hypothesis or prediction, or the effectiveness of that a specified course of action.

Herein, we refer to the confidence, trustworthiness, and acceptability characteristics as *'credibility'*. In M&S, credibility is effectively synonymous with confidence, trustworthiness and acceptability, and are often used interchangeably [9–11]. Relative to M&S testbeds, credibility refers to that attribute of a testbed information and development process that involves the belief of the observer/user. Thus, credibility perception is intrinsically subjective. It is loosely tied to the information about the derivation process of a testbed, so that the reliability of the process only adds to credibility if the observer/user well-understands and appreciates the process and associated limitations [12]. Consequently, to trust the credibility of testbed information and development process, an observe/user must also trust that the testbed authors have the appropriate competency to apply the process, and did so correctly

The testbed constructs in most publications focus on specific sectors or applications but lack detailed and sufficient views about the testbed use and results. In addition, testbed constructs appear characterised by dissimilar approaches to developing and demonstrating security M&S research [13]. These introduce uncertainties and weak arguments for the reliability of each contribution, which is exacerbated by trade-offs between obtaining *'generality'* to a broader set of ICS applications or *'specificity'* to a specific pressing operational process or application problem at hand.

In this work, we interpret credibility as; *how well a security testbed (system, process, and (or) outputs) is able to reflect and advance confidence, trustworthiness and acceptability as a correct representation of a real system* [14], *and suitable to use for explorative cases or scenarios*. This, definition of credibility links to the non-inclusion or non-coverage of certain model design/simulation attributes that can add to a strong reflection of a real system. For ICS security testbeds, this is important as it can impact the accurate resolution of security and safety issues in real ICS. Simulation credibility can suggest how well a system, process, component, and/or outputs reliably re(presents) the actual system. One way of achieving this is through outlining the relevant design considerations or factors that support development and use, and which can support confidence and acceptability. We have not found any work that sufficiently addresses this. This presents another challenge when developing capabilities to support research objectives and when evaluating the quality of a testbed and related research. Thus, it is difficult to make statements or to demonstrate how previous work supports or improves confidence in actual ICS scenarios. It is previously acknowledged that having and following a guiding/benchmarking structure is crucial in proving the relevance and significance of ICS testbeds and associated areas [15].

To address the above challenges and needs, we draw from a systematic review of existing ICS security testbed work to identify relevant design factors that can provide guidance on security testbed development and use. We then propose a novel conceptual relationship map of credibility-supporting design factors (and their associated attributes) and a process implementation flow structure, for ICS security testbeds. *The mapping structure and implementation process should not be construed as strictly sequential. The succession in the map using arrows is intended to show the relationships amongst design credibility factors and sub-attributes*. Broadly, the relationship map can assist testbed developers and decision-makers in determining suitable design factors and approaches peculiar to their requirements. The process implementation flow can assist with a step-by-step guide on how the attributes in the map can be adopted towards a credible ICS testbed implementation. Together, both structures can support establishing and(or) enhancing the credibility of security-related ICS testbed work. The concepts proposed can also be applied to other security testbed development areas such as wireless sensors and computer networks.

The remainder of this paper is structured as follows; Section 2 presents an overview of work related to ICS security testbed simulation. Section 3 describes our research methodology. Section 4 presents the results and analysis of fundamental design considerations identified in the study. Section 5 describes our proposed mapping structure and process for demonstrating ICS security testbed credibility. Section 6 concludes the work.

## 2. Related Work

The testbed publications reviewed considered those that are closely aligned to ICSs and/or cybersecurity including; the Internet of Things (IoT), cyber ranges, and Cyber Physical Production Systems (CPPS).

Candell et al [2] detail three design/case study ICS testbeds with security scenario demonstrations. For the scenarios, design attributes are not structurally defined from the outset. Instead, design details on process descriptions, components, architecture, protocols, modelling approach, and security scenarios appear unstructured across the paper. This makes it difficult to easily and clearly recognise attributes that might come across as potential design requirements. In addition, the work has a limitation in that it fails to consider credibility-supporting factors such as `evaluation modes and outputs' which can be useful in building credible ICS designs/testbeds [16,17]. Not considering evaluation modes demonstrates a lack of corroboration by parties other than the researchers/authors on the quality and credibility of testbeds or related work. Such work is felt to insufficiently support any claimed credibility by the authors.

Gluhak et al [18] performed a technology-based review of IoT experimental testbeds. They focused on Wireless Sensor Networks (WSNs), and examined the effort required when migrating from WSNs to a global networked infrastructure of IoT. They evaluated existing IoT testbeds based on design challenges and functional characteristics including scalability, heterogeneity, repeatability, federation, concurrency, and mobility. These reflect characteristics considered valuable in contributing to testbed design credibility.

Davis and Magrath [5] surveyed cyber ranges and computer network operations (CNO) testbeds from three broad classifications: Modelling and simulation, Ad-hoc or Overlay, and Emulations. One of the key conclusions is that simulation and emulation are the most common approaches for developing security testbeds due to a reduced cost of implementation, flexibility, scalability, and capacity for easy reconfiguration.

In Siaterlis and Genge [19], a comparative study of nine ICS testbeds is presented. An analysis was conducted using a coarse scale (1-3) to rate six key operational characteristics: Fidelity, Repeatability, Measurement accuracy, Safety, Cost effectiveness and Multiple critical infrastructures, and two sub-characteristics: Cyber and Physical. Besides failing to consider other crucial attributes like Scalability, Modularity, and Flexibility, the authors compared their work against others, but failed to provide any clear bench-marking requirements for the quality evaluations.

Holm et al [20] surveyed 30 ICS testbeds proposed for scientific research. Most of the testbeds were designed for vulnerability analysis, test of defence mechanisms, and educational purposes. Pure simulation of ICS components appeared more common than virtualisation and hardware-based approaches. Testbed fidelity was heavily emphasised. However, factors like repeatability and safe execution were not well-addressed by the surveyed testbed articles.

In Salunkhe et al [21], a conceptual design of CPPS testbeds is presented, based on a review of prior testbeds. These were analysed based on their application sectors, i.e., electrical grid, cybersecurity, network and communications, robotics and manufacturing, IoT, Web and Cloud computing, simulation-based, and others. Results showed cybersecurity to be the dominant area of interest.

Design considerations for security testbeds is clearly a topic of interest across relevant communities and stakeholders. However, using relevant requirements as a means to address design credibility, and how this may be achieved, is currently absent in community discussions and literature. While attributes that can pass as design considerations and development factors have been discussed directly or implicitly in several work, they are fragmented. This restricts the ability to identify a broader set of essential requirements or mappings to appropriate functionalities that can support the credibility of ICS security testbed designs and associated research activities. An outline of security testbed design essentials can help to streamline existing concepts and enable a pathway for suggesting a standardised evaluation or benchmarking approach for ICS security M&S testbeds.

## 3. Methodology

We started by identifying relevant research contributions from related work, and then design factors from which attributes can be drawn.

To provide a comprehensive view of the ICS security testbed space, we opted for a systematic review [22]. This began with an unstructured survey involving online Google searches, applying related titles for a period covering 2008 to 2018. This was used to pick relevant keywords. These were then applied to a structured search in the SCOPUS and Web of Science databases for relevant articles (focusing on finding related keywords within each article's title, keywords, or abstract). Both databases were chosen because together they enable access to more resources with strength of a wider coverage and resource concurrency [23]. Keywords used involved Boolean combinations of `ICS' OR `SCADA' AND `Security Testbeds' OR `Testbeds'. Finally, articles were selected if they contained text describing any variant of ICS security-related testbed design architectures, frameworks, or implementation, as research objectives, or as a tool for validating conducted research.

A qualitative study was also made, involving a three-hour focus group workshop comprising 16 participants with ICS security modelling interests. These were drawn from academic, policy-making, and mixed-interest (i.e., combined interests in academia and policy-making) communities. Participants were asked to provide responses to specific questions. Inclusion criteria for participants was that they had experience or interests in designing, using, or regulating contexts related to ICSs M&Ss.

Out of the 16 participants, four self-identified with policy-making, six with academia, and six with mixed-interests. Answers were collected on boards using sticky notes for each interest group. The goal was to obtain a range of views and experiences from stakeholders on the significance of credibility in ICS security testbeds, especially from a research and development perspective. This was broken down into two precise questions: (i) *Is credibility an issue in the development, use, and utility of security testbed models for ICS? If Yes, why?* (ii) *Design considerations that can build/strengthen credibility of ICS security simulation testbeds and the research that uses them.*

A thematic analysis method using Braun & Clarke's six-phase guide [24], was used to examine the data and to derive insights, as shown in Table 1. Data was collected from the sticky note responses of participant groups and analysed based on research questions following a top-down (theoretical) theme approach. This involved combining semantic and latent-level evaluations to identify more specific patterns in group responses, and exploring any underlying ideas and assumptions that may be associated with the themes.

*Table 1: Thematic Analysis Process*

| Research Questions | Steps | Context | Description |
|---|---|---|---|
| RQ1: Is credibility an issue in the development, use, and utility of security testbed models for ICS? If Yes, why?, RQ2: Design considerations that can build/strengthen credibility of ICS security simulation testbeds and the research that use them. | Step 1 | Familiarising with data | Made notes and jotted down early impressions on the value of credibility in ICS testbed M&Ss, and factors that can enhance credibility from participant post-it notes |
| | Step 2 | Generating initial codes | Coded data segments from written responses on post-it notes in order of relevance to RQs 1 and 2. |
| | Step 3 | Searching for themes | Examined codes and combined related codes into a single theme. |
| | Step 4 | Reviewing themes | Revised and grouped themes in terms of fitness to research questions |
| | Step 5 | Defining themes | Refined grouped themes and defined their essence and implication to study |
| | Step 6 | Write-up | Documented the results and interpretations based on research questions. |

## 4. Results

From the systematic review, 77 articles were identified from the queried databases according to their match with applied search parameters. The relevance of each article was considered based on its title and abstract. Duplications were discarded. This left 41 articles found to contain substantial content on ICS security testbed use. These are presented Appendix A. Relevant design attributes that address ICS testbed design and security simulations were drawn from selected literature. The significance of identified attributes on credibility-building were also analysed comparatively with those obtained from thematic analysis of focus group responses.

### 4.1 Credibility-Supporting Design Factors

ICS security testbeds that can influence evidence-based decision-making on security policies and controls typically depend on the degree of conformity to real system that can be assured. This can influence confidence in the testbed to reliably satisfy a specific intended purpose [17].

The composition of an ICS testbed model typically depends on the knowledge underpinning the testbed development, and the expertise of human experimenter to correctly define, design, integrate and configure testbed components, system and scenarios, documentation, evaluation and comprehension of the results of experimental security scenarios. Thus, the quality of an ICS security testbed and the confidence it can instil depends on the rigour of the operational theory underpinning the testbed, and the expertise in properly applying the theories to logical and convincing outcomes [25]. However, the influence of expertise and knowledge in the credibility of security testbed M&S appears not to be well covered by existing research. This is despite that experimenters not only choose components that can integrate well into a model of a planned system, but also set-up and configure the components and process parameters, execute the processes, collect and analyse the results, and interpret outcomes. People involved in security testbed development have to make decisions including; the security contexts and features that are important to be modelled, the appropriate M&S approach to use, the components and scale to adopt, the core characteristics relevant to the contexts adopted, and the level of evaluation necessary to validate the model. They must have, and employ, the requisite knowledge and skills to logically implement the outlined processes. The lack of sufficient expertise in any areas may result in errors in those aspects of the testbed development. This can cause overall quality degradation and cast doubts as to the credibility of the testbed.

Design considerations that are useful are driven by well-defined `usage objectives' [26]. Because of the trade-off between obtaining highly representative systems and their implementation costs, testbed design considerations and decisions on simulation need to be driven by an intended use [27]. 'Design Objectives' have been cited as a relevant and a key context in testbed preparation, which needs to also align with design configurations [20]. Design objectives and configurations need to be articulated well ahead to provide direction and scope for the development process, as well as to support functional validity and credibility. For simulation testbeds, applicable objectives need to be well clarified, since a design setup can be valid for one objective but not for another [14].

The credibility of ICS security testbeds can also be influenced by their 'Architecture Components' [28,29]. This refers to the common functionality coverage in an ICS setup comprising any combinations of the ICS functional areas; (i) Physical Process (PP), (ii) Field Devices (FD), (iii) Communications Gateway (CG), and (iv) Control Centre (CC) [2,3]. Component architecture also covers aspects of communications protocols; consisting of either IP routable and(or) IP non-routable protocols [2]. Incorporating more aspects of the basic architecture components helps to clarify security issues related to specific components and their implications across the entire ICS network. A broader coverage of components within a common architecture can enable the simulation of wider contexts and enable better realism of an ICS from architectural perspectives. These can also support attaining a more holistic expression of security tests, and insights into the entire system impacts. They lend credence to the resulting testbed and the research that uses it.

Testbed reliability can also be supported by demonstrating certain 'Core Operational Characteristics' [28,29] that underpin the structure and operation of a testbed. The core characteristics can take structural and functional dimensions [1,30]. These comprise of behavioural attributes that are expressed in testbed operations such as the ability to; reflect the real nature of a system (fidelity), add or remove components or test scenarios (modularity), and log status of test scenarios (monitoring and logging). These also cover attributes that refer to testbed performance indicators. These include the ability to; easily use the testbed (usability), adapt it to new applications or scenarios (adaptability), and be open to improvements and modifications (scalability). These features are normally off-shoots of functional features [31]. The relevance of these core operational characteristics in supporting testbed credibility has been acknowledged [1,20,30]. Thus, demonstrating these attributes within a simulation testbed design adds some assurances that can advance trustworthiness and acceptability of the testbed and associated research.

The 'Simulation Approach' adopted for a testbed also contributes to its perceived reliability [28,29]. This refers to the structural and procedural formation of the components that constitute a simulation system testbed. Broadly, this can be classified into three: (i) Physical Simulation (PS) – involving purely real infrastructure components, (ii) Semi-Physical Simulations (SPS), sometimes called 'Hardware in the Loop' – involving a combination of real, emulated and/or virtualised abstractions of ICS components (i.e., a mix of Emulation and implementation-based approaches), and (iii) Software-based Simulations (SBS) – involving the simulation of components on a single, purely software platform. Other terms for these categories include real system (hardware and software), computer emulations or virtualisation (including hardware-in-the-loop), pure software-based simulations [20] or live, virtual, and constructive (LVC) simulations [32,33].

*Real or live* simulation involves actual/real world control system components operating on/with actual/real-world ICS set-up and protocols. Despite using real components, this is considered a simulation because cyber-attack processes and scenarios are simulated, and not truly conducted against any live target adversary control system [33]. An example includes using actual operators, actual network devices, actual components, and actual non-emulated/simulated software. *Emulated or virtual* simulation involves actual ICS components interacting with limited or representative ICS system models and vice versa. A *'representative'* simulation model is one that offers the operationally relevant partial or complete interactive interfaces, protocols, and features of the actual component or system. A simulation model is said to be *'limited'* if part of its components does not provide the relevant interactive interfaces, protocols, of the actual component. Examples include; having the emulators of components such as a PLC running on a virtual machine or replaying a logged real-life attack onto virtual or live systems. *Pure software-based or constructive* simulation approach involves the models of 'limited' or 'representative' components interacting with limited or representative system models. A typical example is simulating internet-scale traffic generation and background noise [33].

The choice of M&S approach can be influenced by factors including; the experience or expertise of humans involved [34], the desired degree of representation or capability of an actual system [35], the cost of development, and the budgeted development time [28]. In particular, the expertise of the human developer can affect how, and the degree of detail captured in a simulation testbed. Physical, real or live simulations typically enable the most representation of real system and data fidelity and is more likely to be credible than the other two approaches.

An ICS testbed's *'Evaluation Process'* can also influence design quality and credibility [16]. This refers to the procedures through which assessments are performed to determine how well a testbed's design or related outputs are correct, and(or) acceptable. The purpose of testbed evaluation is to demonstrate with appropriate evidence that a testbed set-up and its scenario results fit the use intended, and do not present any intolerable risks. Having such evidential information in hand can support well-informed and confident decisions throughout a security M&S life cycle [12]. ICS testbed evaluation helps to clarify on the correctness of a testbed set-up, and its usefulness in addressing real-world industrial system needs. To build or enhance credibility, simulation testbeds, scenarios, data, and results all rely on suitable evaluation. This should demonstrate fulfilment of relevant reliability factors including design objectives, structural, behavioural, and performance characteristics in line with intended use. Such evidence can exist in varied degrees, supporting a scale of credibility and acceptance. While evaluation proofs may be offered by testbed authors, they may be better accepted when coming from other sources, e.g., independent reviewers. However, the best evidences of credibility are likely to come from public institutions, standardisation or certification bodies such as National Institute of Standards and Technology (NIST), the UK's National Physical Laboratory (NPL) and The Institute of Engineering and Technology (IET).

The three contexts of testbed M&S evaluation described can be more technically termed as: *verification, validation,* and *accreditation* [36]. Verification describes the process of clarifying that an ICS testbed model implementation and its associated data correctly represent the developer's specifications. Validation defines the process of determining the degree to which an ICS testbed model and its associated data provide a correct representation of the real-world ICS system from the perspective of the intended uses of the testbed. Accreditation describes the official certification that a testbed simulation model or a federation of testbed and its associated data is acceptable for use for a specific purpose [37]. Each of the evaluation category seeks to answer a unique question that captures a specific testbed simulation idea. Verification answers; *Was the testbed built right?* Validation answers; *Was the right testbed built?* Accreditation answers; *Is the built testbed believable enough to be used?* That an evaluation process can transition from verification to validation and then to accreditation reflects an incremental appraisal process, whose results can provide stronger evidence(s) and ground(s) to persuade confidence, belief, trustworthiness, and acceptability of the testbed simulation outputs.

Arguably, persuading credibility in ICS security testbeds and associated research can involve demonstrating that component setups, functional and application approaches, experimental processes and results are clear and sound. Typically, these include maintaining a set of associated documentation including records demonstrating that testbed systems conform to design goals, architecture components, functionality set-ups and applications, experimental scenarios and measurement outcomes, and evaluation procedures, as applicable in a real-world context. The broad set of documentation should cover context breakdowns of testbed process descriptions including process model schematics, protocols, logical architectures (zones and enclaves), physical architectures, control strategies and parameters, module and component descriptions, assembly details, measurement data collections, evaluation metrics (security and operational). Relevant testbed details need to also include the security requirement descriptions following prescribed or guiding security standard, such as the standard series of ISA99, Industrial Automation and Control System Security [38], or similar contexts in NIST 800-82 [4], and NIST Advanced Manufacturing series 200-1 [39]. The intention is to help users/experimenters become familiar with the technologies, test and evaluation processes involved, and to serve as reference manual. Indicating the qualification and expertise of the human experimenter can support confidence and acceptance in the associated research and its outputs. This can also contribute to improved rigour in supporting decision makers' assessments.

### 4.2. Quantitative Review of Literature

This section presents a quantitative analysis of the literature on credibility factors in ICS testbed research. In analysing the semantic description of security design objectives, eight broad themes have emerged. These include: Threat Analysis, Vulnerability Analysis, Attack Analysis, Impact Analysis, Defence Mechanism Test/Analysis, Education and Training, Creation of Policies and(or) Standards, and Performance/Quality of Service Analysis. Their occurrence across the reviewed work is summarised in Table 2. *'Attack Analysis'* (63.41%) and *'Defence Mechanism Tests/Analysis' (*56.1%*)* are the two most common objectives for designing and using ICS security testbeds for research. Other dominant design objectives for ICS security testbeds include: Impact Analysis, Vulnerability Analysis, and Education & Training.

Again, from Table 2, analysis of the adoption of simulation approaches showed that 21.95% of studied work combined Software-based and Semi-physical (emulation, virtualisation, or HIL) to realise the desired ICS security testbeds systems, processes, and tests. 19.51% combined elements of all three methods (i.e., SBS + SPS + PS), and 12.20% combined Semi-physical (emulation, virtualisation, or HIL) with Physical (Hardware or Software) methods. Fewer works were exclusive in their methods with 17.07% each supporting purely physical simulations and purely software-based approach. 12.20% used a form of Semi-physical method – involving either emulation, virtualisation, or HIL techniques, and often only simulated just a part of the ICS architecture, rather than the complete setup or components.

For architecture components coverage, results in Table 3 show that nearly half of the reviewed research defined and(or) adopted design component structures that covered all four broad functional areas of ICS described earlier. Between 3 and 11 works covered three function areas, and on average 2 works covered two function areas. CGs appear to be the function area most explored, with a coverage of 95.12%. This is followed by CC components (90.24%) and PP components (82.93%).

For core characteristics, a total of fourteen distinct key operational characteristics were found across the projects as shown in Table 2. 70.7% of the reviewed work contained statements and descriptions that related to one or more of the fourteen core characteristics. Researchers more commonly focused on fidelity (41.46%) than the other characteristics. There were 26.83% references to *scalability/extensibility*, 8 instances each (19.51% each) of references to *flexibility/adaptability*, and *repeatability/reproducibility* characteristics, with the remaining characteristics having fewer results.

Concerning evaluation processes, results in Table 4 show that more than half (56.10%) of the reviewed projects lacked information relating to any form of evaluation to verify, validate or accredit their works. 19.51% mentioned evaluation approaches that relate to design/scenario comparisons against common standards and reference model documentation, which may imply a validation process by independent parties. Examples of standards referenced include: NIST SP 800-82 R2 [4], PERA Reference Model [40], and IEC 60870-5-104 TCP/IP Communications [41]. 14.63% of works used a verification approach – showing at completion that their works satisfied prescribed design objectives. No work demonstrated any form of accreditation, neither did any project demonstrate evaluation to a level suitable for accreditation.

*Table 2: Analysis of ICS Security-Related Design Objectives*

| Design/Simulation Objectives (Security-Centric) | Percent of Total (%) |
|---|---|
| Attack Analysis | 63.41 |
| Defence Mechanism Tests/Analysis | 56.10 |
| Impact Analysis | 41.46 |
| Vulnerability Analysis | 36.59 |
| Education and Training | 24.39 |
| Threat Analysis | 9.76 |
| Performance/QoS Analysis | 2.44 |
| Creation of Policies and(or) Standards | 2.44 |
| | |
| **Design/Simulation Approach:** | Percent of Total (%) |
| SBS + SPS | 21.95 |
| SBS + SPS + PS | 19.51 |
| PS | 17.07 |
| SBS | 17.07 |
| SPS + PS | 12.20 |
| SPS | 12.20 |
| *Key to notations:*<br>- *Software-Based Simulation = SBS,*<br>- *Semi-Physical Simulation (Emulation or /Virtualisation / HIL) = SPS,*<br>- *Physical Simulation = PS,*<br>*'+' used to reflect combination of approaches.* | |

Table 3: Analysis of Architectural Component Simulation Coverage

| Design/Simulation Coverage: | Percent of Total (%) |
|---|---|
| CC + CG + PP + FD | 46.34 |
| CC + CG + PP | 26.83 |
| PP + CG + FD | 9.76 |
| CC + CG + FD | 7.32 |
| CC + FD | 4.88 |
| CC + CG | 4.88 |
| **Characteristics** | **Percent of Total (%)** |
| Fidelity | 41.46 |
| Scalability or Extensibility | 26.83 |
| Flexibility or Adaptability | 19.51 |
| Reproducibility or Repeatability | 19.51 |
| Modularity | 17.07 |
| Cost-Effectiveness | 9.76 |
| Measurability & Measurement Accuracy | 9.76 |
| Isolation or Safe Execution | 7.32 |
| Usability | 4.88 |
| Diversity | 4.88 |
| Interoperability | 2.44 |
| Monitoring & Logging | 2.44 |
| Openness | 2.44 |
| Complexity | 2.44 |

*Key to notations:*
- *Communications Gateway = CG*
- *Physical Process = PP*
- *Control Centre = CC*
- *Field Device/Components = FD*

*'+' used to reflect combination of component classes covered.*

Table 4: Analysis of Testbed Evaluation Process

| Process Category | Evaluation Method | Percent of Total (%) |
|---|---|---|
| - | Not Mentioned | 56.10 |
| *Verification* | User-defined Requirements | 14.63 |
| *Validation* | Standards and Reference Model | 19.51 |
|  | Prior Works | 4.88 |
|  | Real ICS | 2.44 |
| - | Unreferenced Architecture | 2.44 |

### 4.3 Qualitative Analysis of Participant Workshop

This section presents the results of thematic analysis of the responses and feedback from the focus group workshop.

For the first questions (RQ1) on whether credibility is an issue in the development, use, and utility of security testbed models for ICS, we find the responses from participants to be unanimously affirmative. There is common agreement that demonstrating credibility is crucial in ICS security testbed-related work. It was common view that clear guidelines on how credibility may be built in ICS security M&S are currently lacking, and not being emphasised enough to command attention and response of testbed security experiment developers

To support the opinion for the importance of demonstrating credibility in the development, use, and utility of security testbed models for ICS, participants identified *"building or enhancing trust"*, and *"supporting real-world applications"* as key drivers for engagement. There is a need to trust and accept as reliable, the design, structure and process implementations of an ICS security testbed, the research that uses it, and any associated results. Being able to depend on the testbed to demonstrate the functionalities and processes expected in real system domains is also vital.

The capacity to *'enable analysis'* also resonated as a common theme that was emphasised during focus group discussions. Analysis dimensions highlighted in this regard include: *"behavioural impact analysis"*, *"accident impact analysis"*, and *"modular-based analysis"*. These highlight capability benefits that can be gained from using ICS testbeds for security analysis. Thus, the criticality of ICSs to societal function makes for the need to ensure a significant degree of certainty and accuracy about any analysis context engaged. The mentioned analysis dimensions can also pass as potential ***design objectives*** for a security testbed and were found useful to support the context being studied.

On the second question (RQ2) on the design considerations that can build/strengthen the credibility of ICS security simulation testbeds and the research that use them, disparate feedbacks were aggregated from the three stakeholder categories. Policy Makers identified available institutional resource capability (cash and skills), demonstrating a shared/cascaded development responsibility, design interoperability and flexibility. Interoperability resonated in the points from the academic group alongside 'demonstrating object-oriented scenario setup', 'capturing system layer-based simulation', and 'ability to simulate automated load, failure handling, and decision-making'. The Mixed Interest group also acknowledge the importance of experts (knowledge and skills) along with an ability to replicate real world scenarios.

The responses from participants were aggregated and harmonised into similar themes following the analysis steps outlined in Table 1. The resulting common themes expressed attributes related to testbed M&S *design*, *process*, *structure*, *organisation*, *application*, and *capability*. For example, responses like *"capturing system layer-based simulation"*, *"including computational infrastructure"*, and *"including object-oriented scenario setup"* related to design factors. Concerning capability factors, one response read *"expert opinion is important"*. Responses related to structural factors include: design *"flexibility"*, *"design interoperability"*, and *"design fidelity"*.

### 5. Discussion on Enabling Credibility Factors in ICS Security Testbeds

It is found from the combined review of existing work that the following factors contribute to the trustworthiness of ICS security testbeds and(or) associated research: *(i)* clearly defined security-related design objectives and security scenarios, *(ii)* the type(s) of simulation approach(es) and the degree of abstractions involved, *(iii)* the scope of architecture components covered, *(iv)* the reflection of core characteristics, and *(v)* the testbed evaluation methods. The recurrence of these attributes in literature gives an idea of their relevance too.

For most of the ICS testbed work studied, not involving evaluation to a level that can support accreditation supports the argument that insufficient emphasis has been given to the significance of building credibility. The lack of a form of evaluation characterises works in this area, which is probably influenced by the widespread lack of emphasis in existing standards and best practice guidelines. It seems that researchers and experts do not perceive the need to address such contexts and attributes in their work as necessities to demonstrating quality and stimulating acceptance.

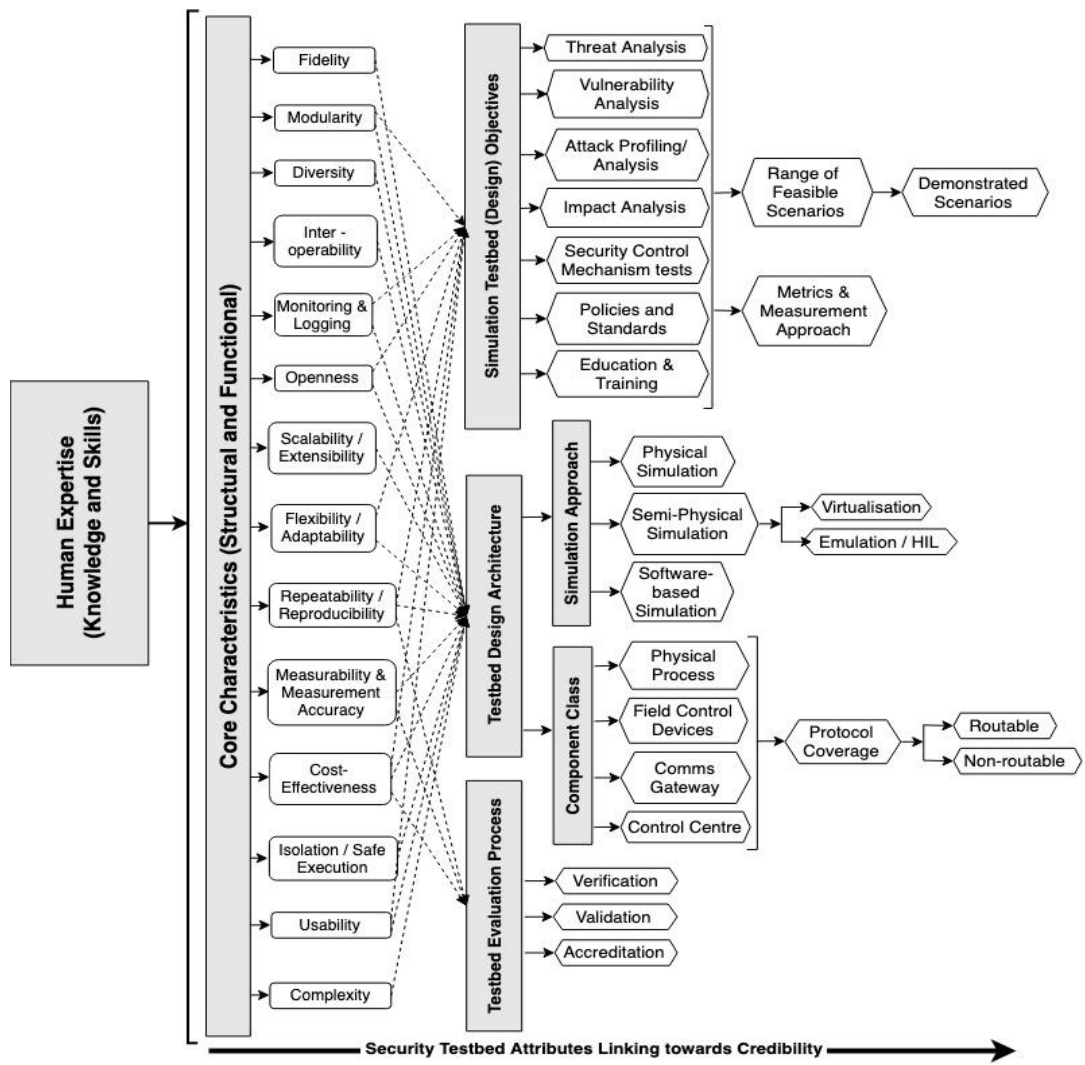

*Figure 1: Mapping Structure for ICS Testbed Credibility Demonstration*

This leaves experimenters short in addressing open questions on how to prove that associated testbeds produce claimed objective parameters and/or valid results, without showing how the verification and/or validation was achieved. Often, since these concerns do not appear to be raised publicly or by industry bodies, they are often swept aside or ignored. Although testbeds tend to have more documentation that serves as a reference guide enabling users to familiarise themselves with the technologies, understanding key contexts that support credibility is crucial to guide researchers and experimenters, since typically user documentation is not of a standard fit for publication. This is perhaps a limitation of existing work, which exposes researchers to the risks of missing valuable information that could underpin credible designs and experiments, particularly those that contribute to repeatability and measurability. A possible solution lies in identifying the factors that are most important to specific sectors and/or applications. This should be explored while considering the trade-off between specificity and generality of a testbed's purpose and following a structured approach in selecting and implementing attributes that can inform the credibility of the setup and/or associated research.

Thematic analysis suggests a strong emphasis on *impact*, evidenced by authors' concerns and focus on demonstrating and learning from negative impacts before they happen, and the quest to achieve resilience. Losses that arise from compromising impacts and the need to reduce or completely avoid system consequences is an associated reason for emphasis. Although significant, these contexts represent just one out of the range of potential design objectives or benefits of the testbed security modelling approach.

There are overlapping views between the themes and codes in thematic analysis and the factors identified from systematic study. Although the terms used to describe contexts appear to vary in both views, the semantics point in a similar direction. Results from the time-constrained focus group are not as detailed and encompassing as those from the systematic study. However, the data available still demonstrate common ideas. For example, response codes under *capability factors* can be linked to *human expertise* in knowledge and skills. Indeed, the opinion of experts depends on their knowledge and skills in the context considered. This can in turn affect the potential credibility level. Responses under *design factors* can be related to both *testbed design objectives*, while combining the responses under *design, process,* and *organisational factors* point to *architectural design attributes*. Responses under *structural factors* can link to *core operational characteristics*. Thus, there is a degree of coherence between the two perspectives concerning the perceived factors that contribute to building or enhancing credibility in ICS testbed security modelling.

### 5.1 Mapping Credibility Characteristics to Security Testbed Configurations

We have developed a mapping structure that outlines the relationships between testbed design factors and show how the identified design factors co-relate to support credibility. As shown in Figure 1, considering these elements and features together can greatly support building a compelling story about ICS security testbed design, setup, and its use and utility for security analysis. Such a narrative can provide a wider understanding of an ICS testbed's composition, functionalities, abstraction, simplifications, assumptions (where available) and test/experimental results, underlining the need for a reliable representation of the real system being modelled or analysed.

i) **Human Knowledge and Expertise**

Clearly, the degree of knowledge and expertise available to the experimenters influences the quality of decisions made concerning the realisation of a security testbed, spanning defining design objectives through to evaluating their implementation.

Resource requirements for security testbed M&S, e.g. experimentation time, budget, and available technology infrastructure, can also influence the parameters of choice, and the level of fidelity achievable. These requirements make it difficult to achieve a generic testbed setup for a span of skills categories, especially for low-skilled users. Documentation that clarifies the context and appropriate level of user expertise can inform confidence in, and the reliability of inferences drawn from experiments. The likelihood is that testbed research from more experienced researchers will potentially provide more depth of analysis and give great confidence in reliability.

The knowledge and experience of testbed developers/users can help to identify core characteristics that need to be captured in specific testbed modelling contexts and scenarios. Once identified, these core characteristics can help guide the characterisation of design objectives and define a range of scenarios to explore. Consequently, relevant metrics and measurement approaches can also be determined. Expertise also informs the appropriate design architecture, components, and simulation methods (including associated sub-attributes). The knowledge and experience of experimenters also contributes to the level of evaluation that can be undertaken.

As a minimum, modern ICS Security testbed experimenters require expertise in both ICS and IT systems development, together with an ability to: adopt appropriate modelling approach, techniques and tools, configure test applications, execute test scenarios, collect and interpret results. These steps require experience as they are susceptible to errors arising from insufficient knowledge and skills. One way is to engage external professional expertise in areas where researchers have limited experience and aptitude, by employing experts in ICS technology development to handle ICS operations design and implementation, while security researchers focus on the security-related experimentation.

ii) **Core characteristics (Structural and Functional)**

We believe that the core characteristics outlined are important as they individually contribute to measures that help establish or advance credibility. It seems that there are characteristics that contribute more than others to overall credibility. The ranking of importance varies across functionalities and application domains, and often based on targeted objectives. However, results from analysis and occurrence frequencies of characteristics can provide suggestions on how the relevance of these characteristics is viewed by the security simulation design community. The number also provides a way of potentially ranking characteristics. For example, the requirement for demonstrating the *fidelity* of a simulation testbed and/or its use seem of greatest significance. This is apparently followed (in order) by *scalability (extensibility), flexibility (adaptability or controllability), repeatability (reproducibility), modularity, measurability/measurement accuracy, cost-effectiveness, safe execution/isolation, diversity, and usability*. To build credibility, it is important for ICS testbed system and associated research to consider and demonstrate these characteristics. Evidencing as many as possible of these characteristics improves the confidence of decision-makers and other stakeholders to consider and accept the results of testbed designs, thereby improving their value.

*Fidelity* refers to the degree of correlation between security simulation or test predictions and real world observations [15,20,42,43]. It quantifies the degree of representativeness between a testbed setup and an actual system, in terms of tools (hardware and software technologies), functionalities and tasks. The degree of fidelity can typically be determined by the simulation approach adopted – either software-based simulation (SBS), semi-physical simulation (SPS), physical simulation (PS), or combinations of these. Physical simulation is typically considered to have the highest fidelity while purely software-based simulations, the least [32].

*Scalability or Extensibility* refers to the characteristic to grow the size of a testbed setup (network) and functionality [15,32,44]. This can be demonstrated by the ability to add or migrate components (e.g., sensors & actuators) to existing operational testbed subsystems, thereby increasing capabilities or functionality without significant re-organisation or re-design. Examples of how this may be achieved are demonstrated thus; for software-based simulation approaches using *SciLab simulators* to add Field Devices; for semi-physical simulation techniques using *Virtual Machines* to emulate Control Centre components; and for physical simulations – using *real subsystems such PLCs* as Field Devices [15,43].

*Flexibility or Adaptability* describes the ability to easily and swiftly re-define and repurpose a simulation system and setup for alternative use cases [3,32,45]. It can also be viewed as *'controllability'* – emphasising the ability and extent to enable the control of environment variables. This can be theoretically expressed in design/simulation objectives and practically demonstrated in design/simulation architecture. For example, an ability to show that a simulation system initially purposed for security vulnerability analysis can be easily re-structured to perform security impact analysis.

*Repeatability or Reproducibility* refers to the characteristic whereby similar outputs/outcomes are obtained from identically replicated designs/testbed setups. Exact copies of designs and testbed setups or security test scenarios should produce identical or statistically consistent results [20,46]. One way this characteristic can be obtained is through full documentation of design and process configurations, as well as security scenarios [15,43,46]. Other researchers can thereby obtain consistent results by applying the same configurations to directly recreate and test scenarios.

*Modularity* describes a design capability that allows easy adaption to changing requirements, including complexities and flexibility in industrial operations [44,47,48]. It involves developing ICS testbed structures that can accommodate continuous improvements. It is typified by a design that can accommodate real components, emulated nodes, and network simulators (data traffic) such as the OPNET modeller, which can enable a typical system-in-the-loop (SITL) capability [49–51]. Such design concepts and provisions can improve system understanding, reduce complexity, increase flexibility, and facilitate the reuse of components [52]. Implementing modularity can provide a structured approach and an action path that realises, through validated module re-use, incremental credibility of a security simulation testbed with respect to environment, data, and results.

*Cost-effectiveness* is a property that relates to achieving testbed design objectives and scenarios within financial budgets that are affordable for research purposes [20,53]. The emphasis is on using smaller budgets/costs (in comparison to actual system costs) to achieve the same design objectives and scenarios (including architectural setups and configurations) as the real. This can be achieved through setups that simulate numerous components and services consolidated into a single portable testbed system [51]. For example, using virtual machines and other virtual infrastructures to emulate control workstations, servers and other ICS components [51,54], which can result in a cost-efficient alternative to using real and proprietary hardware workstation and server systems. This is typically subjective and varies across projects, depending on budgetary availabilities and research/test requirements. Often, a trade-off and balance is required between the cost of constructing testbeds and the fidelity of the system, and the decision that needs to be informed [53].

*Measurability* and *Measurement accuracy* describes the ability to ensure that the process of testing or replicating cyber security scenarios via testbeds can be quantified, and that such measurements do not interfere with corresponding outputs [20,42]. This can be demonstrated by including tools (e.g. sensors) or features for verifying attributes like traffic flows and response times amongst components. The capability to show and document data and values associated with these features also needs to be demonstrated [19,53] typically at the documentation stages of an evaluation procedure (e.g. verification).

*Safe execution or Isolation* of tests describes a characteristic that ensures cyber security scenarios and activities are performed in a secure and isolated approach and environment, such that they do not increase risk or impact safety in the real environment [55]. This can easily be demonstrated using network segmentation approaches [56] to separate plant networks from enterprise networks and processes. The use of access control policies at various network layers is another approach typically implemented at communication gateway components such as access point devices [55].

*Usability* refers to the ability for a testbed to be readily employed by reasonably skilled operators, with little likelihood of simulation misuse [20,32]. This is essential to cope with different skill sets of potential users. Usability can be demonstrated through adopting design and developing structures using components that enable friendly user interfaces [57]. For example, using virtualisation and VLANs to enable the easy integration of testbed components in the CG section of ICS architectures [15,43].

*Diversity* refers to the ability of an ICS testbed to incorporate a varied range of components without undermining the capacity for scalability as discussed earlier. An effective testbed need to be able to mirror a variety of ICS setups [15,43,58]. This includes demonstrating where feasible and necessary; market-driven heterogeneity in components (vendor products e.g., Siemens, Allen Bradley, Schneider), protocols (e.g., TCP, UDP, OPC, Modbus/TCP, DNP3, EthernetIP) and processes (e.g., manufacturing, assembly, traffic control, water treatment) employed in a testbed setup. This can provide valuable ICS security insights from legacy, contemporary, and future outlooks, and deployments that reflect industrial practices, enabling a variety of experimental setups and scenarios. These can help advance system credibility for practical applications [3].

*Interoperability* refers to the ability of combinations of SBS, SPS, and PS testbed simulation components to interface, communicate, exchange and use information to achieve desired objectives. This can typically be demonstrated in the development of hybrid ICS testbeds and security experiments involving PS components such as Control workstations that connect to IED interfaced with SPS techniques like virtual machine servers and virtual communication components [51].

*Monitoring & Logging* describes the ability to observe and record process execution and to optimise event logging for security purposes [15,43]. One way of achieving this is through implementing a measurement enclave with Syslog tools and traffic monitoring systems to keep track of operational activities [3]. This can be better achieved through automated granular data flows – understanding data sources and pathways to help resolve undesirable impacts on process functionality [59].

*Complexity and Openness* describe two related attributes identified as valuable in modern ICS testbed designs. *Openness* defines the capability of a testbed simulation setup to support remote access or data openness [15,43]. While *complexity* ensures that architectures are represented in a transparent manner such that a single point of data access or extraction can be enabled from different network zones or segments of the ICS [15,43,58].

As shown in Figure 3, the mapping can be quite complex with a range of one-to-many connections that may not be clear at first glance. Contextual description may be needed to clarify the specific attributes involved as presented in section *5.1*. Most (14 arrows) of the characteristics appear to map to the *'simulation design architecture'* factor and associated attributes, compared with *'design objectives'* and *'testbed evaluation process'* which had fewer (7 and 3 arrows respectively) mappings. This suggests that the most significant task and proportion of effort for establishing credibility in ICS security testbeds and associated research lies in simulation design architecture. This is exemplified by the simulation approach adopted and the design components and functionalities (hardware and software) employed. The design architecture needs to be carefully considered to ensure capture of the necessary credibility-supporting characteristics and backed with sufficient evaluation processes to maximise system trustworthiness as being representative of the real world.

iii) **Design Objectives**

Design objectives are a vital consideration as pointed out in 4.1, since ICS security testbed design architecture, attributes and decisions must be driven by usage intentions. A majority of ICS security-related research activities seem to focus on investigating cyber-attack feasibilities, the capability of security controls and defence mechanisms, and the analysis of attack impacts, instanced by successful attacks or failed security controls.

However, it is also important to understand the existence and nature of vulnerabilities in industrial control systems and components. Often this builds on the assumption that vulnerabilities nearly always exist in ICSs, since they are by default presumed to lack security. This means that experimenters typically focus on understanding random attack modes, often through penetration testing, and on ascertaining robustness against specific attacks given certain security measures applied. Along these ideas, there is a risk of losing sight of the susceptibilities that may emerge due to system complexity, interdependencies, and cascading impacts. Yet these are the types of insights that are needed to support better security decision-making, and that should be considered in emerging and future testbed security analysis work. Formulating system-level and organisational security policies and standards is another security testbed design priority that requires more attention in order to re-focus the technical community – from attributing greater relevance to tasks related to establishing security than those of security governance and standardisation.

In addition, clearly defining testbed design objectives serves to resolve the typical tension between the high cost of deploying security testbed components, and the degree of similitude to the real system. For example, a testbed to determine vulnerabilities in ICS sub-systems like PLCs may not require the significant implementation and representation effort of an entire industrial architecture set-up, which can be expensive and unnecessary. Model approaches that involve combining the target component/module with virtual/software-based components can also be considered. Both measures can significantly reduce the cost of testbed development. Articulating specific high-level security-related design objective(s) well ahead of implementation can help with the decisions related to the control choice(s) to be made. For example, Fovino et al [56] described the analysis of cyber-attacks and impacts as objectives for their testbed-related security study. They further narrowed these objectives to encompass "SCADA system phishing with DNS poisoning, DoS Worm, and Modbus/DNP3 protocol worm". Similarly, Bergman et al [55] described their objective to aim at supporting the analysis of security control and impact. They clarified this further by indicating that in the context of the objective mentioned, their work explored *"testing the impacts of network segmentation and SSL encryption using OpenVPN"*. Providing this level of specificity proved necessary to support a clearer understanding of what the work entailed in scope and design.

### iv) Simulation Approach

For the simulation approach, all three schemas – real/live, emulated/virtual, and software-based/constructive, appear to have significant support in the user community. However, combining multiple simulation approaches seems more popular than using each alone. Besides aiming for a greater similitude to real system, another motivation involves exploring combinations that enable one approach to cover for the limitations of another. Often, the choice of method(s) to combine is influenced by the degree of fidelity desired, the cost and affordability of development involved, and the time available for testbed development. The requirement for researcher/user expertise is a cross-cutting theme. Achieving high realism involves using infrastructures that work in the real environment; these are often expensive. Also, longer periods may be required to complete set-ups and configure the system and test processes. Often, decisions need to be weighted by trade-offs between attributes based on the defined objectives.

### v) Architecture Components

From results, the coverage of communication gateway (CG) components seems to dominate other architecture functional groups. This may be due to the mature nature of research and development in digital system/network communications gateway infrastructure, which increases its popularity over other functionality groups. The coverage frequency of component classes may also be driven by the degree of class criticality in testbed considerations.

Often, a broader coverage of functional areas, component classes and applicable routing protocols in a testbed architecture can depend on the objective(s) and scope of desired test scenario(s). Recapping the earlier scenario of assessing the vulnerabilities of a single FD device, this is unlikely to require building an entire SCADA system network but could involve a more direct approach of executing security audits on the desired device without necessarily embedding it in an operational ICS network. The hardware-in-the-loop (HIL) approach is a good technique to use. Another example involves structures that include CG components such as routers and switches where routable protocols are required. However, when tests do not involve the flow of data over router-based sub-systems and connections, then non-routable networks and protocols are appropriate. It may be safer to consider routable protocols (e.g., Modbus TCP and DNP3) in ICS security testbed designs since they are by design better adapted to secure configuration across network paths than the non-routable protocols (e.g., DeviceNet), which are better placed to handle perimeter-based security. Notwithstanding the choice, incorporating all four functionality classes into an ICS security testbed design can allow for wider contexts to be mirrored, which typically provides a better representation of an ICS from both architectural and operational perspectives.

### vi) Design Evaluation Process

The existence and rigour of evaluation process affects the credibility of ICS security simulation testbeds. The three levels of evaluation – *verification, validation,* and *accreditation (VV&A)* – indicate the possibility of a scale of credibility.

*Verification* enables project experimenters to double-check that adopted/defined security design/simulation requirement(s), system functions and processes, and any associated data, correctly represent the experimenter's conceptual description and specifications. It spans a range of contexts such as; *(i) verifying testbed structure and defined objectives* – justifying that testbed attributes are identified and usage objectives are clearly defined with sufficient accuracy, *(ii) verifying design problems* – proving that the problem adopted contains the actual environment problem, and is sufficiently well-formed to allow sufficiently credible solutions to be obtained, *(iii) verifying functions* – activities that demonstrate that testbed system functions and operations accurately mirror known real system behaviours relative to defined objectives, (iv) *verifying solutions* – activities demonstrating that the outputs and results reflect the known outcomes in real systems subject to the same parameters and operational conditions [25]. *Verification* is performed by the security testbed experimenters – a way of self-corroboration – to support credibility by demonstrating that predefined requirements in design, functionality, and outputs are well-satisfied. Techniques that can be applied for this can include: Desk checking, model review, result analysis, instrumentation-based testing, functionality testing, and sensitivity analysis [60].

*Validation* often comes from parties non-affiliated to the context being validated and seeks to establish the extent to which an ICS security testbed and associated data, mirror the real-world intended use scenario. Typically, it may involve the work of third-parties in repeating processes and methodologies of verification, to ensure agreement between the observed or known behaviour of real system components and processes with that of the testbed simulation system. It also includes ascertaining whether any difference between the two are acceptable given the testbed's intended use. This crucially helps to timely identify and rectify issues and avoid unnecessary misuse of evaluation time and other resources by users. *Validation* can be done using different methods, including based on historical data, through comparison with other testbed simulators, from expert judgements, parameter sensitivity analysis, or predictions [61].

*Accreditation* aims to certify that a testbed set-up and/or associated data are acceptable for a defined purpose. An accreditation evaluation leads to a recommendation that a certifying official or body authorises that a specific testbed M&S set-up or tool can be used for the purpose it is designed [62]. Thus, an accreditation often relies on the evidence(s) from verification and validation, and from additional in-depth and multi-level evaluations to establish more concrete attestation by an official certification body (typically government-based) following an independent assessment. In testbed model accreditation, the acceptability criteria are identified well ahead, and then the evidential knowledge from verification and validation processes is applied to ascertain how the intended use of the testbed model is impacted [63]. This way, accreditation not only focuses on the intended use, but also on the requirements for adopting testbed models.

Not all testbed M&S structures require an evaluation to accreditation level, so this should be pursued only if necessary. We believe that the level of evaluation necessary to support credibility and adoption can depend on factors such as project costs, available time and resources. In a resource-constrained setting, the cost of an evaluation process that involves VV&A activities on ICS security testbeds M&S can be prohibitively high compared to what is available or affordable. The unavailability of appropriate reference data, information from development products, documentation of past evaluations, can increase evaluation costs. Thus, it is advisable that the extent to which evaluation investments are made be weighed against potential risks of reaching; weak conviction, a bad decision, and huge scale of adverse impact; stemming from unreliable testbed M&S results or uncertainty in the evaluation process and life cycle. For example, although accreditation can afford greater credibility, it can take longer, and extend project schedules to achieve the in-depth documentation and assessments that may be required by certification bodies. This can be costly to achieve and may be less of an issue in the case of validations or verification – with the possible consequence of impacting perceived credibility of relevant testbeds or related work.

Unarguably, it is beneficial to pursue transitioning from verification to validation, and finally accreditation, as this provides a stronger evidence base and grounding for credibility. However, testbed developers and decision-makers responsible for evaluation investments need to consider risk possibilities related to defects in testbed set-ups, simulation scenarios and processes, hardware components, software elements, data, or even misjudged testbed capabilities by users. The level of tolerance considered acceptable should be part of the determining factors for the level of evaluation to reach.

## 5.2 Credibility-building Process

Besides understanding the crucial design factors that can help to improve credibility of testbed simulation systems and their relationship as shown in Figure 1, it is crucial to adopt a structured approach in applying the factors into design implementation processes. The process can guide ICS security testbed experimenters on the steps to follow in their M&S process to persuade confidence and acceptability. This is shown in Figure 2.

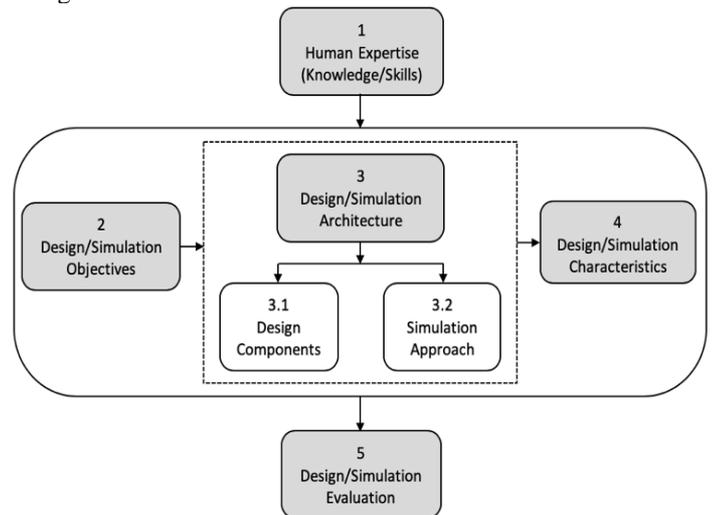

*Figure 2: ICS Testbed Credibility-building Process*

The first step in the credibility-building process should start with human expertise. Typically, based on the expertise of the developers involved, it is essential to have a testbed development plan – covering aspects of ICS and the IT and how they are integrated. Expertise in these areas could be demonstrated and inferred from the quality of context, design setup, and descriptions provided, which should reflect the real system. Testbed security design and simulation objectives need to be clearly defined and described in context to set the target scope.

The second step in the process involves defining the security design objective of interest. This should describe in clear context the security purposes for which the testbed design and scenarios are being implemented. The objectives can be singular, or a range of target security capacities or intents that could be achieved with the design. Expert knowledge and skills, coupled with clearly defined design/simulation objectives, contribute to defining an appropriate architecture.

The third step involves defining the design/simulation architecture. The architecture will encompass hardware, software, and protocol components and sub-systems. It also includes the simulation approach adopted, or any combinations thereof which are relevant for the intended test strategy and cope. These are aimed at persuading a high degree of confidence, trustworthiness acceptance of outcomes – credibility, as shown in Figure 1.

The fourth steps involve identifying and demonstrating core design/simulation characteristics that have been deemed relevant and considering any trade-offs. This gives better idea on the context and scope of security modelling and simulation adopted.

Once initial design and security test(s) have been completed, it is crucial to engage a final step of evaluating security scenario outcomes against initial characteristics intended in the plan. Evaluation must also include checking that the setup/simulation outcomes satisfy target objective(s), design architectures, and core functional characteristics. Evaluation can be done by various agents; from the experiment developers to external individuals or standards organisations, in order to corroborate initial claims on the reliability of the security testbed and/or results.

## 5.3 Summary

The drivers that underpin design considerations in existing research indicate clear tensions between the '*generality to suit a broad set of ICS domains and/or applications*' and '*specificity to solve impending simulation challenges peculiar to domain or application*'. Common interests tend to focus on the latter (i.e., specificity).

The *generality* attribute can enable a capacity to mirror contexts and applications pertaining to multiple ICS domains without the need for significant re-configuration. Often this leads to a downside resulting from a lack of depth in system and process replication and analysis. Only high-level views of the system are captured, making up for 'breadth', but lacking in 'depth' of context coverage. *Specificity* to particular security modelling simulation problems enables the adoption of design attributes which favour more focused coverage and in-depth analysis down to a detailed level. This enables a more tailored and a better understanding of simulation systems behaviour and performance. The approach appears more common in the community of designers and users perhaps because of the less demanding requirements related to; engaging a narrower area and view, and a lower level of expertise and specialisation.

Despite the underlying use of similar hardware and software infrastructure in various ICS domains, design architectures, protocols, functionalities, operations often vary amongst sectors. Each sector application usually involves complex component and process interactions that require specialised knowledge and skills to implement. It is rarely feasible to find expertise in depth that spans multiple infrastructures and that would enable a broader ICS testbed implementation that is fit for multiple purposes. This would enable in-depth modelling and simulation of multi-modal sector applications, architectures, components, protocols, processes, interactions and complexities. Specialisation seem to occur because developers and users engaged in ICS security testbed design and simulation typically have expertise (deep knowledge and skills) in a specific ICS domain or a narrow area of application.

## 6. Conclusion.

Several factors need to be considered when evaluating the reliability of simulation systems, the research that use them and their outcomes to guide the perception of credibility. Our research has explored literature and interacted with stakeholders to identify relevant factors that can provide guidance on ICS security testbed development and use, and which can support the decision on testbed credibility. We developed mapping framework (see Figure 1) outlining testbed design factors and how they co-relate to support credibility. This is used by following a testbed credibility-building process (see Figure 2) which provide a structured approach to apply the factors into design implementation.

Demonstrating credibility in ICS security simulation testbeds remains an issue of concern, and the requirements to support this need to be streamlined. Building or enhancing credibility typically arises mainly from architectural coverage, characterised by the adopted implementation approach, selected components, and the demonstration of a reasonable degree of evaluation. These need to be engaged through a structured process, from defining security testbed design/simulation objectives to evaluating the work using the most feasible/available approaches. ICS security researchers and developers must strive to achieve fundamental architectures that are representative of real-world systems and can allow appropriate, yet realistic testing.

The expertise (knowledge and experience) of researchers and developers is crucial relative to achieving defined objectives and scenarios. Clear security-related design objectives defined from the outset can help drive the testbed development process, maintain a focused direction, and contribute reliability to the outcome. Clarifying the testbed simulation approach provides a path to understanding the tools and techniques adopted and their simulation capabilities. It also provides the information needed to reproduce and validate simulation testbed designs/systems and associated research. A clear outline of the architectural composition, and the adopted testbed simulation approach, increases the potential for demonstrating scientific rigour and repeatability, adding credibility to claims of quality and fidelity. Demonstrating evaluation procedures across verification, validation, and(or) accreditation can help attest to the satisfaction of quality, value, and acknowledgement in communities beyond the immediate designers, developers, and researchers. Including evaluation details can help resolve queries related to if and how a security testbed was validated and persuade a wider acceptance of a claimed credibility state. Having simulation systems and testbeds subjected to this type of multi-level evaluation process against available credibility criteria, can evidence quality and trustworthiness for critical decision-making.

It is beneficial to capture the core characteristics within a testbed setup. However, choosing the most important compliance characteristics within a specific project will depend on the project's core objectives and scope. Trade-offs may be needed, and considering the available resources/capabilities, certain characteristics may be incorporated or maximised at the expense of others. New attributes can also be considered based on emerging interests and evolving dynamics in the system or context of application. The proposed relationship mapping approach can promote effective and well-organized procurement of systems and sub-system components guided by clearly defined design requirements; responding to system and functional dynamics, and the endorsement of the relevant community of stakeholders. It can thereby streamline the task of setting requirements and reduce the costs of both infrastructure development and sub-system integration. It can lead to greater consistency and efficiency in developing research related to ICS security testbeds, building on what already exists. Most conveniently, by combining this with the growing trend and capability for federating ICS security testbeds, as has been keenly advocated and explored in recent publications, the potential is increased for testbed availability and interoperability. Furthermore, a federation architecture/system can minimise the diversity in design structures between different and physically dispersed testbed infrastructures. For future work, we will explore how ICS testbeds are evaluated, and how credibility is tested.

# Appendix A: ICS Security-related Testbed Works

| | Authors | Paper Title | Institution | Country | Objectives | Approach | Landscape/Coverage | Credibility Requirements | Evaluation/Validation |
|---|---|---|---|---|---|---|---|---|---|
| 1 | Giani et al 2008 | A Testbed for Secure and Robust SCADA Systems | UC Berkeley | USA | VA, DMT | SBS, EM, PS | PP, FD, CG, CC | Not Mentioned | Not Mentioned |
| 2 | Hieb et al 2008 | Security Enhancements for Distributed Control Systems | University of Louisville | USA | DMT | SBS | CC/PP | Not Mentioned | Not Mentioned |
| 3 | Queiroz et al, 2009 | Building a SCADA Security Testbed | *RMIT University* | Australia | AA | SBS, EM | PP, CG, CC | Modularity, Fidelity | Base on Prior works |
| 4 | Bergman et al 2009 | The Virtual Power System Testbed and Inter-Testbed Integration | University of Illinois at Urbana-Champaign | USA | IA, DMT | SBS, EMU | FD, CG, CC | Isolation, Reproducibility, Scalability, Flexibility, Fidelity | Not Mentioned |
| 5 | Kush et al 2010 | Smart Grid Test Bed Design and Implementation. | Queensland University of Technology | Australia | VA, TA, IA | Virtualisation, | PP, FD, CG, CC | Flexibility, Extensibility (Scalability) | Comparison with User-defined functional requirements |
| 6 | Chunlei et al 2010 | A Simulation Environment for SCADA Security Analysis and Assessment | Tsinghua University of Beijing | China | AA | SBS, EM, PS | PP, FD, CG, CC | Extensibility, Adaptability (Flexibility) | Unreferenced SCADA Reference Architecture |
| 7 | Fovino et al 2010 | An Experimental Platform for Assessing SCADA Vulnerabilities and Countermeasures in Power Plants | European Commission Joint Research Centre | Italy | AA, IA | PS | PP, FD, CG, CC | Repeatability, Safe Execution | Not Mentioned |
| 8 | Hahn et al 2010 | Development of the PowerCyber SCADA Security Testbed | Iowa State University | USA | ET, AA, | EM | PP, CG, CC | Fidelity | Based on NERC & NIST Requirements |
| 9 | Stefanov and Liu, 2011 | Cyber–Power System Security in a Smart Grid Environment | University College Dublin | Ireland | AA, TA, VA, IA | SBS | PP, CG, CC | Not Mentioned | Not Mentioned |
| 10 | Dondossola and Garrone, 2011 | Cyber Risk Assessment of Power Control Systems – A Metrics weighed by Attack Experiments | Ricerca sul Sistema Energetico | Italy | AA, IA, DMT | SBS | PP, FD, CG, CC | Not Mentioned | Compliant with design standards (IEC 60870-5-104 TCP/IP Communications) |
| 11 | Morris et al 2011 | A control system testbed to validate critical infrastructure protection concepts | Mississippi State University | USA | ET, VA, DMT | PS | PP, FD, CG, CC | Fidelity | Not Mentioned |
| 12 | Mallouhi et al 2011 | A Testbed for Analyzing Security of SCADA Control Systems (TASSCS) | University of Arizona | USA | DMT | SBS | FD, CG, CC | Not Mentioned | Not Mentioned |
| 13 | Jin et al 2011 | An Event Buffer Flooding Attack In DNP3 Controlled Scada Systems | University of Illinois at Urbana-Champaign | USA | AA, VA | PS | CG, CC | Flexibility, Extensibility (Scalability) | Based on a Real Design Testbed |
| 14 | Almalawi et al 2013 | SCADAVT–A Framework for SCADA Security Testbed Based on Virtualization Technology | RMIT University | Australia | AA, IA, | Virtualisation, | PP, FD, CG, CC | Usability, Scalability, Fidelity, Modularity | Not Mentioned |
| 15 | Sayegh et al 2013 | Internal Security Attacks on SCADA Systems | American University of Beirut | Lebanon | AA, VA | PS | PP, CG, CC | Not Mentioned | User-defined requirements |
| 16 | Shahzad et al 2013 | Secure Cryptography Testbed Implementation for SCADA Protocols Security | University Kuala Lumpur | Malaysia | DMT | SBS | Not Mentioned | Not Mentioned | Not Mentioned |
| 17 | Urias et al 2013 | Supervisory Command and Data Acquisition (SCADA) system Cyber Security Analysis using a Live, Virtual, and Constructive (LVC) Testbed | Sandia National Laboratories | USA | VA, AA | SBS, EM/V, PS | PP, CC, CG | Modularity, Interoperability, Scalability, Cost-Effectiveness, Fidelity | User-defined requirements |

| | Authors | Paper Title | Institution | Country | Objectives | Approach | Landscape /Coverage | Credibility Requirements | Evaluation/Validation |
|---|---|---|---|---|---|---|---|---|---|
| 18 | Stites et al 2013 | Smart Grid Security Educational Training with Thunder Cloud: A Virtual Security Test Bed | Tennessee Technological University | USA | ET, VA, AA | EM/V | PP, CC, CG | Cost-Effective | user-defined requirements |
| 19 | Hahn et al 2013 | Cyber-Physical Security Testbeds: Architecture, Application, and Evaluation for Smart Grid | Iowa State University | USA | VA, IA, AA | SBS, EM, PS | PP, FD, CG, CC | Scalability, Modularity, Extensibility, Fidelity (Accuracy) | Not Mentioned |
| 20 | Gao et al 2014 | An Industrial Control System Testbed Based On Emulation, Physical Devices And Simulation | Technical Assessment Research Lab | China | VA, DMT, | SBS, EM, PS | PP, FD, CG, CC | Fidelity, Modularity, Repeatability, Measurability, Cost-Effective | compliant with ANSI/ISA-99 standard |
| 21 | McLaughlin et al 2014 | Multi-attribute SCADA-Specific Intrusion Detection System for Power Networks | Queen's University Belfast | Ireland | AA, DMT | SBS | PP, CG, CC | Fidelity | User-defined requirements |
| 22 | Genge and Siaterlis, 2014 | Cyber-Physical Testbeds - EPIC | European Commission Joint Research Centre | Italy | AA, IA, DMT, Network QoS Effects on cyber attacks | SBS, EM | PP, FD, CG, CC | Fidelity, Measurement Accuracy, Repeatability, Scalability, Safe Execution (Safety) | Compliant with design standards (IEEE 9, 30, 39 and 118) |
| 23 | Haney and Papa 2014 | A framework for the design and deployment of a SCADA honeynet | The University of Tulsa | USA | DMT, | SBS, EM/V, PS | | Not Mentioned | Not Mentioned |
| 24 | Candell et al 2015 | An Industrial Control System Cybersecurity Performance Testbed | National Institute of Standards and Technology (NIST) | USA | DMT, IA | SBS, EMU/HIL, PS | PP, FD, CG, CC | Diversity, Flexibility, Scalability, Fidelity, Security Analysis, Extensibility | Compliant with NIST SP 800-82 Security guidelines |
| 25 | Singh et al 2015 | A Testbed for SCADA Cyber Security and Intrusion Detection | Centre for Development of Advanced Computing | India | DMT, AA, | SBS, EM | PP, FD, CG | Not Mentioned | Not Mentioned |
| 26 | Koutsandria et al 2015 | A Real-Time Testbed Environment for Cyber-Physical Security on the Power Grid | Sapienza University of Rome, Arizona State University and Lawrence Berkeley National Laboratory | Italy and USA | AA | SBS, EM, PS | PP, FD, CG, CC | Fidelity, Repeatability | Not Mentioned |
| 27 | Farooqui et al 2015 | Cyber Security Backdrop: A SCADA Testbed | National University of Sciences and Technology | Pakistan | AA, IA | SBS | PP, CG, CC | Flexibility, Usability | Not Mentioned |
| 28 | Jarmakiewicz et al 2015 | Development of Cyber Security Testbed for Critical Infrastructure | Military University of Technology | Poland | DMT, | EM, PS | PP, CG, CC | Fidelity | Compliant with Standard |
| 29 | Ghassempour et al 2015 | A Hardware-in-the-Loop SCADA Testbed | University of South Florida | USA | AA, DMT | SBS, EM/HIL, | CC, CG | Not Mentioned | Complaint with IEEE-C37.118 and Modbus protocol design |
| 30 | Ghaleb et al 2016 | SCADA-SST: A SCADA Security Testbed | King Fahd University of Petroleum & Minerals | Saudi Arabia | AA, DMT, IA | SBS/EM | PP, FD, CG, CC | Modularity, Extensibility, Reproducibility | |
| 31 | Hink and Goseva-Popstojanova, 2016 | Characterization of Cyberattacks aimed at Integrated Industrial Control and Enterprise Systems: A case study | West Virginia University | USA | ET, VA, DMT, TA, IA, | PS | PP, CC, CG | Believed To Be Representative Of The Real Systems | Validation |
| 32 | Cruz et al 2016 | A Cybersecurity Detection Framework for Supervisory Control and Data Acquisition Systems | University of Coimbra | Portugal | DMT, TA, | SBS, EM | PP, FD, CG | Fidelity, Repeatability, Data Accuracy | Validated by comparing normal and attack scenarios results |

| | Authors | Paper Title | Institution | Country | Objectives | Approach | Landscape/Coverage | Credibility Requirements | Evaluation/Validation |
|---|---|---|---|---|---|---|---|---|---|
| 33 | Mathur and Tippenhauer, 2016 | SWaT: A Water Treatment Testbed for Research and Training on ICS Security | Singapore University of Technology and Design | Singapore | ET, IM, DMT, | PS | PP, FD, CG, IoT, HHC | Third Party Developers | none |
| 34 | Korkmaz et al 2016 | ICS Security Testbed with Delay Attack Case Study | Binghamton University | USA | IA, DMT, ET | PS | PP, FD, CG | Fully Consistent With Industry Instrumentation Standard | Not Mentioned |
| 35 | Ahmed et al 2016 | A SCADA System Testbed for Cybersecurity and Forensic Research and Pedagogy | University of New Orleans | USA | AA, ET, DMT | PS | PP, FD, CG, CC | Fidelity, Modularity | Not Mentioned |
| 36 | Neg et al 2016 | A SCADA testbed for Cyber Security Education & Research | Indian Institute of Technology | India | ET, AA, VA, DMT, IA | EM, PS | PP, FD, CG, CC | Flexibility | Not Mentioned |
| 37 | Alves et al 2016 | Virtualization of Industrial Control System Testbeds for Cybersecurity | University of Alabama in Huntsville | USA | AA, IA | EM/V, PS | PP, FD, CG, CC | Fidelity, Measurement Accuracy, | Not Mentioned |
| 38 | Soupionis et al 2016 | Cyber Security Impact on Power Grid Including Nuclear Plant | European Commission, Joint Research Centre (JRC) | Italy | AA, VA | SBS, EM/HIL, | PP, FD, CG, CC | Not Mentioned | Not Mentioned |
| 39 | Green et al 2017 | Pains, Gains and PLCs: Ten Lessons from Building an Industrial Control Systems Testbed for Security Research | Lancaster University | England | ET, VA, AA, DMT, IA | EM, PS | PP, FD, CG, CC | Scalability, Diversity, Flexibility, Fidelity, Monitoring, Logging, Openness, Usability, Complexity | PERA Reference Model & Other prior testbed infrastructures |
| 40 | Koganti et al, 2017 | A Virtual Testbed for Security Management of Industrial Control Systems | University of Idaho | USA | VA, AA, IA, | SBS, EM | PP, FD, CG | Not Mentioned | Not Mentioned |
| 41 | Rubio-Hernan et al 2017 | Security of Cyber-Physical Systems From Theory to Testbeds and Validation | Université Paris-Saclay | France | AA, ET | PS, EM | PP, CG, CC | Repeatability, Cost-Effective | Not Mentioned |